\documentstyle[12pt]{article}
\begin{document}
\newcommand{\newc}{\newcommand}
 
\newc{\be}{\begin{equation}}
\newc{\ee}{\end{equation}}
\newc{\ba}{\begin{eqnarray}}
\newc{\ea}{\end{eqnarray}}
\newc{\bea}{\begin{eqnarray}}
\newc{\eea}{\end{eqnarray}}
\newc{\D}{\partial}
\newc{\ie}{{\it i.e.} }
\newc{\eg}{{\it e.g.} }
\newc{\etc}{{\it etc.} }
\newc{\etal}{{\it et al.} } 
\newc{\ra}{\rightarrow}
\newc{\lra}{\leftrightarrow}
\newc{\no}{Nielsen-Olesen }
\newc{\lsim}{\buildrel{<}\over{\sim}}
\newc{\gsim}{\buildrel{>}\over{\sim}}
 
\begin{titlepage}
\begin{center}
June 1997\hfill       
CRETE-97/16 \\
\vskip 0.5cm
 
{\large \bf
A New Statistic for the Detection of Long Strings in Microwave 
Background Maps.}

\vskip .5in
{\large Talk presented at the Particle Physics and the Early Universe Conference
at the University of Cambridge, 7-11 April 1997.}\\[.05in]

\vskip .5in
{\large Leandros Perivolaropoulos}\footnote{E-mail address:
leandros@physics.uch.gr},\\[.05in]
 
{\em Department of Physics\\
University of Crete\\
71003 Heraklion, Greece
}
\end{center}

\vskip .1in

\begin{abstract} 
\noindent
A  new   statistic  is  briefly reviewed,
designed  to  detect   isolated   coherent   step-like
discontinuities  produced by cosmic strings present at
late  times.  As a  background  I  superpose  a  scale
invariant  Gaussian random field which could have been
produced  by a  superposition  of seeds on all  scales
and/or  by  inflationary  quantum   fluctuations. 
The statistical  variable considered is the
Sample   Mean   Difference (SMD)   between   large
neighbouring   sectors  of  CMB  maps,   separated  by
lines in two  dimensional  maps and points in
one dimensional  maps.  I find that the SMD statistics
can detect {\it at the  $1\sigma$  level} the presense
of a long  string  with  $G\mu (v_s  \gamma_s)={1\over
{8\pi}}({{\delta  T}\over  T})_{rms} \simeq 0.5 \times
10^{-7}$ while more  conventional  statistics like the
skewness  or the  kurtosis  require  a value of $G\mu$
almost an order of magnitude larger for  detectability
at a comparable level.  \end{abstract}
\end{titlepage}
\section{Introduction }

The  purpose  of  this  talk  is  to  briefly review  a new
statistic \cite{p97}  which is optimized  to detect the
large  scale  non-Gaussian  coherence  induced by late
long strings on CMB maps.  The statistical variable to
use  is  the  Sample  Mean   Diffrence   that  is  the
difference of the mean  ${{\delta  T}\over T}$ between
two large  neighbouring  regions  of CMB maps.  I will
first discuss the main  predictions  of models
for  CMB   fluctuations.  Models  based  on  inflation
predict  generically the existence of scale  invariant
CMB fluctuations with Gaussian statistics which emerge
as a superposition  of plane waves with random phases.
On the other hand in models  based on  defects  (for 
pedagogic   reviews  see  e.g.  \cite{b92,p94}),   CMB
fluctuations are produced by a superposition  of seeds
and   are   scale    invariant \cite{p93a,a96}    
but    non-Gaussian \cite{p93b,p93c,g96,g90,mpb94}.
Observations  have  indicated  that  the  spectrum  of
fluctuations  is scale invariant \cite{s92} on scales larger than
about $2^\circ$, the recombination  scale, while there
seem to be  Doppler  peaks on  smaller  scales.  These
results  are  consistent  with   predictions  of  both
inflation  \cite{b94}  and defect models (\cite{p93a,a96}
and  references  therein)  even though  there has been
some  debate  about the model  dependence  of  Doppler
peaks in the case of defects (see the contributions of
N. Turok and of R. Durrer in these proceedings).

Inflation  also predicts  Gaussian  statistics  in CMB
maps for both  large and small  scales  and this is in
agreement with  Gaussianity  tests made on large scale
data so far.  On small  angular  scale maps  where the
number  of   superposed   seeds  per  pixel  is  small,
topological   defect   models   predict   non-Gaussian
statistics.  This   non-Gaussianity   however  depends
sensitively on both, the details of the defect network
\cite{as90,a96}
at the  time  of  recombination  $t_{rec}$  and on the
physical  processes  taking place at  $t_{rec}$.  
On the other hand, the large number of superposed seeds
on large angular scales leads by the Central Limit 
Theorem to Gaussian statistics for angular resolution
larger than $2^\circ$. This statement however, ignores
the large scale coherence induced by large scale seeds
present at late times.
This
large scale coherence can induce specific non-Gaussian
features even on large  angular  scales.  The question
of   how   Gaussian   are   the   topological   defect
fluctuations on large angular scales will be the focus
of this talk.

The reason that the defect induced fluctuations appear
Gaussian  in maps with large  resolution  angle is the
large number of seeds  superposed on each pixel of the
map.  This, by the Central Limit  Theorem, leads to a
Gaussian probability  distribution for the
fluctuations ${{\delta  T}\over T}$. Non-Gaussianity
can manifest itself on small angular scales comparable
to the minimum correlation length between the seeds.

These   arguments   have  led  most  efforts  for  the
detection of defect  induced  non-Gaussianity  towards
CMB maps with  resolution  angle less than  $1^\circ$
\cite{g90,fm97,p93a}.
There is however a loophole in these  arguments. They
ignore the large scale coherence induced by the latest
seeds.  Such large  scale  seeds must exist due to the
scale  invariance  and they  induce  certain  types of
large  scale  coherence  in CMB maps.  This  coherence
manifests itself as a special type of  non-Gaussianity
which  can be  picked up only by  specially  optimized
statistical   tests.  Thus  a   defect   induced   CMB
fluctuation  pattern can be decomposed  in two parts.
A  Gaussian   contribution   $({{\delta  T}\over  T})_g$
produced mainly by the superposition of seeds on small
scales and possibly by inflationary fluctuations and a
coherent contribution  $({{\delta T}\over T})_c$ induced
by the  latest  seeds.  The  question  that we want to
address  is:  What  is the  minimum  ratio  $({{{\delta
T}\over  T})_c}\over  ({{{\delta  T}\over  T})_g}$ of the
last seed  contribution on ${{\delta  T}\over T}$ over
the  corresponding   Gaussian   contribution  that  is
detectable at the $1\sigma$ to $2\sigma$ level.

\section{Cosmic Strings}

I will  focus on the case of cosmic  strings.  In this
case the  contribution of the latest long string comes
in   the   form   of   a    step-like    discontinuity
\cite{ks84,g85}  coherent on large  angular  scales.
As a toy model we may first consider a one dimensional
pixel array of standardized,  scale invariant Gaussian
fluctuations    with    a    superposed    temperature
discontinuity of amplitude $2\alpha$ \cite{p97}.

A  statistical   variable  designed  to  pick  up  the
presence  of this step is the  Sample  Mean Difference 
(SMD) $Y_k$ 
which assigns to each pixel of the map the  difference
of the mean of pixels $1...k$ minus the mean ${{\delta
T}\over  T}$ of pixels $k+1 ...  n$.  
It is straightforward to show \cite{p97} that
\ba 
Y_k &=& \Delta{\bar  X}_k + 2  \alpha  {{n-i_0}  \over  {n-k}}
\hspace{1cm}  k\in  [1,i_0]  \\ Y_k  &=&  \Delta
{\bar X}_k + 2 \alpha {i_0 \over k} \hspace{1cm}
k\in [i_0  ,n-1] 
\ea
where $k$ labels the $k^{th}$ out of the $n$ random
variables of the pixel map and $i_0$ is the location of
the superposed coherent discontinuity. $\Delta{\bar  X}_k$ is the SMD of the underlying scale invariant Gaussian background.
The SMD  average
statistic  Z is defined as the  average  of $Y_k$ over
all k  
\be    Z={1\over    {n-1}}
\sum_{k=1}^{n-1}  Y_k \ee

It is  straightforward  to show  that the mean
${\bar Z}$ over many realizations and locations of the
step function is ${\bar Z}=\alpha$ and the variance of
${\bar Z}$ depends  both on the number of pixels n and
on  the   step   function   amplitude   $\alpha$

\be 
\sigma_Z^2  = {{2\ln  n} \over n} +  
{1\over 3}\alpha^2   
\ee

The 
condition  for  detectability  of  the  coherent  step
discontinuity at $1\sigma$ level is that ${\bar Z}$ is
larger  than  the  standard  deviation  of  $Z$  which
implies  that  $\alpha  > 0.2$ for  $n\simeq  O(10^3)$
where  $\alpha$  is measured in units of the  standard
deviation  of  the  underlying  Gaussian  map.  It  is
straightforward  to apply a similar  analysis  for the
more  conventional  statistics  skewness and kurtosis.
That analysis \cite{p97} shows that the minimum value of $\alpha$
detectable at the $1\sigma$ level is about an order of
magnitude  larger.  It is  therefore  clear  that  SMD
statistical  variable  is  particularly  effective  in
detecting    coherent    step-like     discontinuities
superposed on Gaussian CMB maps.

A detailed  understanding of the  effectiveness of the
SMD   statistic   requires  the  use  of   Monte-Carlo
simulations.  In  order  to  verify   the   analytical
results for the mean and  variance of the SMD variable
I first  applied  this  statistic  on  one-dimensional
Monte-Carlo   maps   of   scale   invariant   Gaussian
fluctuations   with  step  function   superposed.  The
results  were in good  agreement  with the  analytical
predictions shown above and are described in detail in
\cite{p97}. 
Here  I   will   only   discuss   the   two
dimensional Monte Carlo simulations.

\begin{figure}
\begin{center}
\unitlength1cm
\begin{picture}(6,4)
\put(-3.5,-1.5){\includegraphics{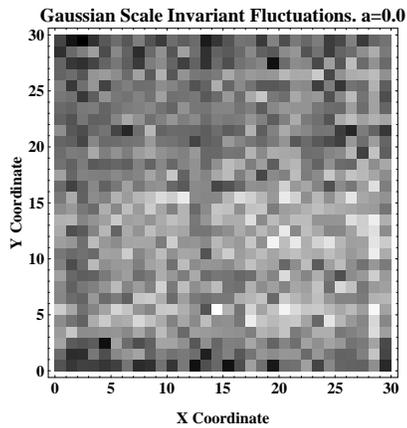}}
\end{picture}
\end{center}
\caption{ A standardized two dimensional pixel array of scale invariant Gaussian 
fluctuations. No step function has been superposed.}
\end{figure}


Figures 1 and 2  show
$30 \times  30$ pixel  maps of  standardised  Gaussian
scale invariant  fluctuations without (Fig. 1) and with (Fig. 2)
a coherent step function superposed.  The amplitude of
the  superposed  coherent  seed  is  $\alpha  =  0.5$.

\begin{figure}
\begin{center}
\unitlength1cm
\begin{picture}(6,4)
\put(-3.0,-1.7){\includegraphics{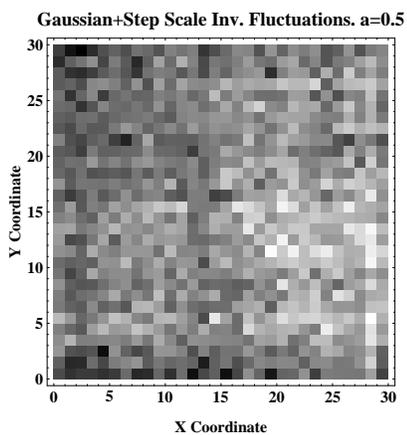}}
\end{picture}
\end{center}
\caption{The two dimesnsional array of Figure 2 with a superposed coherent 
step-discontinuity of amplitude
$\alpha= 0.5$ defined by the random points $(x_1,y_1)=(13.6,18.1)$ and $(x_2,y_2) = (9.4,20.4)$}
\end{figure}

%
Uncorrelated  noise has also been included  with noise
to   signal   ratio  of  0.5.  The   scale   invariant
background  $X(i,j)$ was  constructed in the usual way
by taking its Fourier  transform  $g(k_1,k_2)$ to be a
Gaussian complex random variable.  Its phase was taken
to be  random  with a  uniform  distribution  and  its
magnitude was a Gaussian  random  variable with 0 mean
and  variance  equal  to  a  scale   invariant   power
spectrum.  The SMD was  obtained by randomly  dividing
the map in two sectors  and taking the  difference  of
the  means of the two  sectors.  The SMD  average  was
then obtained by averaging  over many randomly  chosen
divisions for each map realization.  Using 50 such map
realizations  I  obtained  the mean  and the  standard
deviation of the  statistics  skewness and SMD average
for several values of $\alpha$.  The results are shown
on  Table  1  and  indicate   that  the   statistics
skewness  and  kurtosis  can not  identify  a coherent
discontinuity  of  amplitude  $\alpha  < 1$ but  would
require   a   much   larger    amplitude    for   such
identification.  

\vspace{1cm}

{\bf  Table 1}:  A    comparison    of   the
effectiveness  of the  statistics  considered in
detecting  the  presence  of  a  coherent   step
discontinuity with amplitude  $2\alpha$ relative
to the  standard  deviation  of  the  underlying
Gaussian  map. The SMD
average was obtained  after  ignoring 150 pixels
on each  boundary  of the Monte Carlo maps.  The
discontinuities  were also  excluded  from these
300  pixels.  This  significantly  improved  the
sensitivity   of  the  SMD  test.  
\vskip  0.2cm
\begin{tabular}{|c|c|c|c|}\hline  {\bf  $\alpha$
}&{\bf  Skewness  }&{\bf  Kurtosis   }&{\bf  SMD
Average } \\ \hline 0.00 &$0.01 \pm 0.10$ &$2.96
\pm 0.15$ &$0.01 \pm 0.24$\\  \hline 0.25 &$0.01
\pm 0.09$  &$2.95 \pm 0.15$  &$0.28 \pm 0.26$ \\
\hline  0.50  &$0.02 \pm 0.14$  &$2.94 \pm 0.18$
&$0.63 \pm 0.29$ \\ \hline 1.00 &$0.03 \pm 0.30$
&$2.78  \pm 0.30$  &$1.21  \pm  0.46$ \\  \hline
\end{tabular} \vspace{4mm}

On the other  hand the SMD  statistic
can identify a coherent discontinuity at the $1\sigma$
to $2\sigma$ level with $\alpha = 0.5$.  For ${{\delta
T}\over  T}_{rms}  \simeq G\mu v_s  \gamma_s > 4\times
10^{-7}$  where  $\mu$ is the mass per unit  length of
the string,  $v_s$ is its velocity and  $\gamma_s$  is
the relativistic Lorenz factor.

The main  points I wanted  to stress  in this talk are
the following:  
\begin{itemize} 
\item The detection of
non-Gaussianity  induced  by cosmic  strings  on large
angular  scales is possible.  
\item For this a special
statistic  is needed  optimized  to pick up the  large
scale  structure  of the latest  seeds.  
\item  Such a
statistic is the average of the Sample Mean Difference
which  can  pick  up  non-Gaussian  features  of  long
strings with $G\mu v_s \gamma_s > 4\times  10^{-7}$ in
maps  with  about  $10^3$  pixels  and with a noise to
signal  ratio of about  0.5 in a  Gaussian  background
with ${{\delta T}\over T}_{rms} \simeq 10^{-5}$.
\end{itemize}

\section*{Acknowledgements}

I wish to  thank Anne Davis  for useful discussions
and the DAMTP of Cambridge University for hospitality 
and support
during the period when this work was in progress. This
work  was   supported by  the E.U. under the HCM 
programs ($CHRX-CT94-0423$, $CHRX-CT93-0340$ and 
$CHRX-CT94-0621$)
as well as by the Greek General Secretariat of Research 
and Technology grants $95E\Delta 1759$ and
$\Pi ENE\Delta$ 1170/95.

\end{document}